\documentclass{PoS}

\usepackage{tabularx}

\title{The DPHEP Study Group:\\ Data Preservation in High Energy Physics}

\ShortTitle{The DPHEP Study Group: Data Preservation in High Energy Physics}

\author{\speaker{David M. South}\thanks{on behalf of the DPHEP Study Group.}\\
  Deutsches Elektronen Synchrotron\\
  Notkestrasse 85\\
  22607 Hamburg, Germany\\
  E-mail: \email{david.south@desy.de}}

\abstract{
  An inter-experimental study group, DPHEP, was formed in 2009 to systematically investigate
  the technical and organisational aspects of data preservation and long-term analysis in
  high-energy physics, a subject which had hitherto lacked clarity in the field.
  The study group includes representation from all major high-energy physics collider-based
  experiments and laboratories, as well as computing centres and funding agencies.
  A major report was released in May 2012, greatly expanding on the ideas contained in
  a preliminary publication three years earlier, and providing a more solid set of recommendations,
  not only concerning data preservation and its implementation in high-energy physics, but also
  the future direction and organisational model of the study group.
  A brief description of the DPHEP Study Group and some of the key messages from the major report
  are presented.

}

\FullConference{36th International Conference on High Energy Physics,\\
		July 4-11, 2012\\
		Melbourne, Australia}

\begin{document}

%%%%%%%%%%%%%%%%%%%%%%%%%%%%%%%%%%%%%%%%%%%%%%%%%%%%%%%%%%%%
\section{Data preservation in the collider era of experimental particle physics} 
\label{sec:dpinhep}

The last 50 years of high-energy particle physics have produced a wide variety of results from many,
often very different experiments.
The energy and intensity frontiers have been probed with increasingly complex accelerator
installations, where new experiments typically, but not always, supersede previous ones with
similar physics programmes.
This period has also witnessed a growth in size of the necessary international collaborations,
as well as the diversity of the data management.
%
%Today, the age of the LHC has truly arrived, and future projects such as a Super-B
%factory~\cite{Aushev:2010bq}, a new lepton-proton collider such as the
%LHeC~\cite{AbelleiraFernandez:2012cc}, a high energy linear $e^+e^-$ collider~\cite{ilc,clic} or
%LHC luminosity upgrades~\cite{hllhc} are at various stages of planning or implementation.
%
Today, the age of the LHC has truly arrived, and future projects such as a Super-B factory, a new
lepton-proton collider such as the LHeC, a high-energy linear $e^+e^-$ collider or LHC luminosity
upgrades are at various stages of planning or implementation.

%%%

After the collisions have stopped and the final analyses are completed, what to do with
the data?
Such a question has probably been posed fairly often in the last 50 years, but until recently
there was no clear policy on this in the HEP community, and it is likely that older HEP experiments
have in fact simply lost the data.
Data preservation, including long term access, has generally not been part of the planning,
software design or budget of a particle physics experiment and so far, HEP data preservation
initiatives have been in the main not planned by the original collaboration, but have rather
been due to the effort of a few remaining knowledgeable people (the preservation of the
JADE $e^+e^-$ data is a well known example~\cite{Bethke:2010ai}).

%%%

The handling of modern HEP data involves large scale traffic, storage as well as migration
and the increasing scale of the data distribution, whilst facilitating analysis, may complicate
any preservation effort.
Where the responsibility lies - with the experiments or the computer centres - is often a grey
area, and the problem of older, unreliable hardware resulting in unreadable disks or tapes
can occur after only 2-3 years.
Additionally, it is important to realise that the conservation of tapes is not equal to data preservation.
The software necessary for accessing the data, which constitutes a major part of the problem,
is usually under the control of the experiments, where key resources, both in funding and person-power
expertise, tend to decrease once the data taking stops.

%%%

A key ingredient to the question of data preservation in high-energy physics is motivation:
why do this at all? 
Can the relevant physics cases be made? This is a rather crucial question, as it is physics where any
justification must have its roots.
And then is the benefit worth the required cost and effort?
To answer these and further questions and address the issue in a systematic way, an international
study group on data preservation and long term analysis in high-energy physics, DPHEP~\cite{dphep},
was formed at the end of 2008.

\begin{figure}[t]
  \begin{center}
    \includegraphics[width=\textwidth, clip=true, trim=2.5cm 5.5cm 7.5cm 6.2cm]{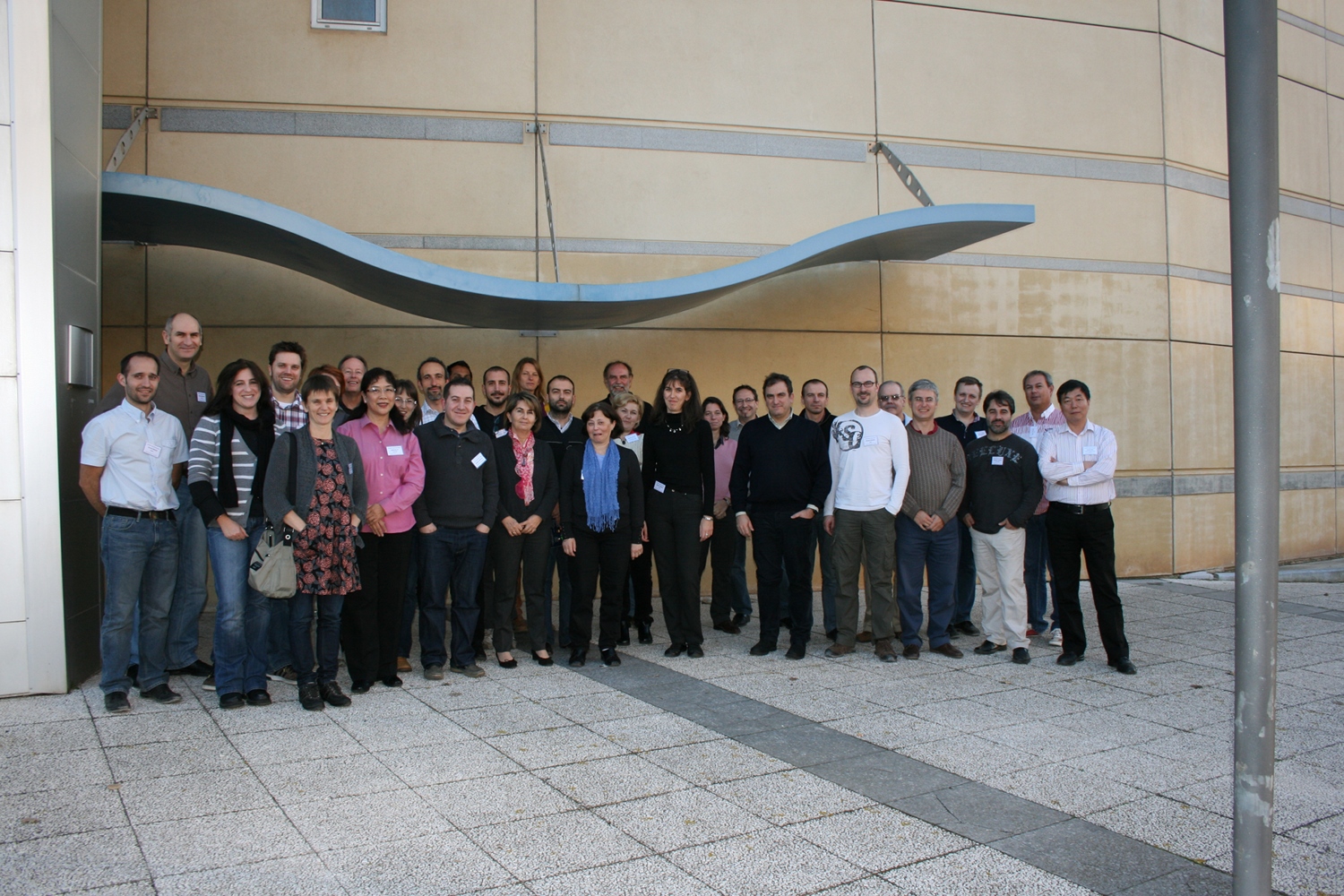}
  \end{center}
  \vspace{-0.4cm}
  \caption{Participants of the sixth and most recent DPHEP workshop, which was held at the Centre de Physique
    des Particules de Marseille in November 2012.}
 \vspace{-0.2cm}
 \label{fig:people}
\end{figure}

The aims of the study group include to confront the data models, clarify the concepts,
set a common language, investigate the technical aspects, and to compare with other fields such as
astrophysics and those handling large data sets.
The experiments BaBar, Belle, BES-III, CLAS, CLEO, CDF, D{\O}, H1, HERMES and ZEUS are
represented in DPHEP; the LHC experiments ALICE, ATLAS, CMS and LHCb joined the study
group in 2011.
The LEP experiments are also visible within the DPHEP effort.
The associated computing centres at CERN (Switzerland/France), DESY (Germany),
Fermilab (USA), IHEP (China), JLAB (USA), KEK (Japan) and SLAC (USA) are all also
represented in DPHEP.  
 
%%%

A series of six workshops have taken place over the last three years, and since 2009
DPHEP is officially endorsed with a mandate by the International Committee for Future
Accelerators, ICFA.
The initial findings of the study group were summarised in a short interim
report in December 2009~\cite{dpheppub1} and a full status report was released in
May 2012~\cite{dpheppub2}, to coincide with the International Conference on Computing
in High Energy and Nuclear Physics (CHEP) 2012~\cite{chep}.
The report contains: a tour of data preservation activities in other fields; an expanded
description of the physics case; a guide to defining and establishing data preservation
principles; updates from the experiments and joint projects, as well as person-power
estimates for these and future projects; the proposed next steps to fully establish
DPHEP in the field.
The remainder of these proceedings briefly summarises the physics case for
data preservation, the various preservation models established and the future working
direction of DPHEP; further details can be found in the latest publication~\cite{dpheppub2}.

%%%%%%%%%%%%%%%%%%%%%%%%%%%%%%%%%%%%%%%%%%%%%%%%%%%%%%%%%%%%
\section{The physics case for data preservation}

When building the physics case for data preservation, four main categories have
been identified:
Firstly, data preservation is beneficial to the long-term completion and extension
of the physics programme of an experiment.
In the case of the LEP experiments a considerable tail exists in the publication rate,
which continues today, and a similar trend is predicted by the HERA experiments and
BaBar.
Up to $10$\% of papers are finalised in the post-collisions period, and prolonging
the availability of the data may result in a gain in scientific output of an experiment.

%%%

Secondly, cross-collaboration and combination of data from multiple experiments may
provide new scientific results, with improved precision and increased sensitivity.
This may occur during the active lifetime of similar experiments at one facility, such as
those at LEP, HERA, or the Tevatron, but may also occur later across larger boundaries,
such as combinations of Belle and BaBar or Tevatron and LHC data.
The preservation of such data would facilitate the comparison of complementary
physics results as well as allowing the independent verification of experimental
observations.

%%%

Thirdly, it may be useful to revisit old measurements or perform new ones with older data.
Access to newly developed analysis techniques as well as the possibility to perform
comparisons to state-of-the-art theoretical models may produce improved or even
new physics results.
Furthermore, unique data sets are available in terms of initial state particles or centre
of mass energy or both, such as the PETRA $e^+e^-$, HERA $e^{\pm}p$ and Tevatron
$p\bar{p}$ collision data, as well as data from a variety of fixed target experiments.
More recently, the early LHC data, taken at centre of mass energies of $900$~GeV and
$2.36$~TeV, as well as the low pile-up $7$~TeV data taken in 2010 also provide unique
opportunities.

%%%

Finally, the value of using real HEP data for scientific training, education and outreach
cannot be understated.
Providing a wide variety of HEP data sets for such analysis, with a corresponding wide
variety of associated exercises and teaching programmes, is clearly beneficial in 
attracting a new generation of inquisitive minds to the field.

%%%

A more detailed description of the physics case for data preservation can be found
in the DPHEP status report~\cite{dpheppub2}, which includes specific examples of
analyses using older data from among others LEP and PETRA, in addition to a description
of the potential of the data from the experiments in the final analysis phase: the
B-factories, HERA and the Tevatron.

%%%%%%%%%%%%%%%%%%%%%%%%%%%%%%%%%%%%%%%%%%%%%%%%%%%%%%%%%%%%
\section{Models of data preservation in high-energy physics}

In order to develop a solid definition of models of data preservation, it is first important
to ask the question: what is HEP data?
The digital information, the data themselves, are clearly crucial, but at least for the
pre-LHC experiments volume estimates for preservation are of the order of a few to
$10$~PB, which is certainly within the storage capabilities of today's HEP based
computing centres.
The range in data volume to be preserved is often a result not only of different sized
data sets, but different types of data: from the basic level raw data, through reconstructed
data, up to the analysis level ntuples.

%%%

However, if one thing may be learned from previous enterprises, it is that the conservation
of tapes is not equivalent to data preservation, and that providing not only the hardware to
access the data but also the software and environment to understand the data are the
necessary and more challenging aspects.
Therefore, in addition to the data the various software, such as simulation, reconstruction
and analysis software need to be considered.
If the experimental software is not available the possibility to study new observables or to
incorporate new reconstruction algorithms, detector simulations or event generators is lost.
Without a well defined and understood software environment the scientific potential of the data
may be limited.

%%%

Just as important are the various types of documentation, covering all facets of an experiment.
This includes the scientific publications in journals and online databases such as
INSPIRE~\cite{inspire} and arXiv, published masters, diploma and Ph.D. theses, as well as a
myriad of internal documentation in manuals, internal notes, slides, wikis, news-groups and so on.
Detailed information about analyses may only be available in internal notes, which may not be
easily available, electronically or otherwise.
Many types of internal meta-data may also exist, such as the details of the detector layout and
performance, hardware replacements, manuals or the documentation of meetings. 

%%%

Finally, the often unique expertise and contributions of collaboration members must also be
considered as another component of HEP data, where particular care is needed to ensure
that crucial know-how does not disappear with losses in person-power, which is liable to happen
towards the end of an experiment.

\begin{table}[t]
\begin{center}
 \renewcommand{\arraystretch}{1.2} 
\begin{tabular}{|l|p{7cm}|p{6.6cm}|}
\hline
\multicolumn{2}{|l|}{Preservation Model}  & Use Case \\
\hline                                        
\hline
1 & Provide additional documentation &  Publication-related information search\\
\hline
2 & Preserve the data in a simplified format & Outreach, simple training analyses and data exchange\\
\hline
3 & Preserve the analysis level software and the data format & Full scientific analysis possible,  based on the existing reconstruction\\
\hline
4 & Preserve the reconstruction and simulation software as well as the basic level data & Retain the full flexibility and potential of the experimental data\\
\hline
\end{tabular}
\end{center}
\vspace{-0.4cm}
\caption{Data preservation levels defined by the DPHEP Study Group.} 
\vspace{-0.2cm}
\label{tab:levels}
\end{table} 

Considering this inclusive definition of HEP data, a series of data preservation levels has been established
by the DPHEP study group, as summarised in table~\ref{tab:levels}.
The levels are organised in order of increasing benefit, which comes with increasing complexity and cost.
Each level is associated with use cases, and the preservation model adopted by an experiment should 
reflect the level of analysis expected to be available in the future.
The four levels represent three different areas, which represent complementary initiatives:
documentation (level 1), outreach and simplified formats (level 2) and technical preservation
projects (levels 3 and 4). 

%%%

The many types of documentation available are described in the previous section, and all
collaborations actively involved in DPHEP pursue level 1 type projects.
Within DPHEP and the member collaborations there are generic ideas, such as common formats 
and user interfaces for outreach.
Such formats are typically based on ROOT, containing particle 4-vectors and basic
event information and can be used for composite-particle reconstruction, finding signals
and other simple exercises.
Simplified data formats also provide an ideal way of transferring data between experiments
and theory groups, allowing new models to be tested.

%%%

Levels 3 and 4 are really the main focus of the data preservation effort, to maintain usable access to
analysis level data, MC and the analysis level software, in addition (in the case of level 4) to the
reconstruction and simulation software.
This re-emphasises the salient point concerning data preservation: it's not about the data,
but about still being able to analyse it.
This may be realised using two alternative paradigms: either keep the current environment alive as
long as possible or adapt and validate the code against future changes as they happen.
These two complimentary approaches are taken by BaBar at SLAC and the HERA experiments at DESY,
both employing virtualisation techniques, but in different ways, as described in detail in
the DPHEP publication~\cite{dpheppub2}.

%%%

More generally, it is worth noting that the implementation of a data preservation model as early as
possible during the lifetime of an experiment may greatly increase the chance that the data will be
available in the long term, and may also simplify the data analysis in the final years of the collaboration. 
Planning a transition of the collaboration structure to something more suited to a long-term
organisation also makes it easier to address issues such as authorship, supervision and access.

%%%%%%%%%%%%%%%%%%%%%%%%%%%%%%%%%%%%%%%%%%%%%%%%%%%%%%%%%%%%
\section{Summary and future working directions}

\begin{figure}[t]
  \begin{center}
    \includegraphics[width=0.695\textwidth]{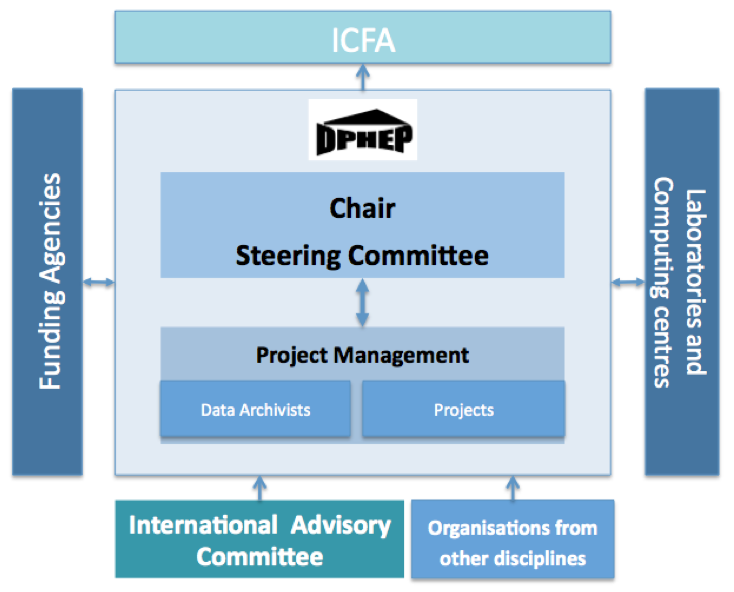}
  \end{center}
  \vspace{-0.55cm}
  \caption{The new DPHEP Collaboration structure and its associations.}
  \vspace{-0.15cm}
\label{fig:dpheporg}
\end{figure}

The DPHEP Study Group has established itself in the HEP community and has reached
a milestone in the publication of its latest report, which contains a comprehensive
appraisal of data preservation in HEP.
The group will continue to investigate and take action in areas of coordination,
preservation standards and technologies, as well as expanding the experimental reach
and inter-disciplinary cooperation, including the full deployment of the various
experiment and laboratory based projects.
In order to do this, the study group will move to a new organisational model, which is
illustrated in figure~\ref{fig:dpheporg}, and make the transition to the DPHEP Collaboration
in 2013, following the recent appointment of the Project Manager position at CERN.

%%%%%%%%%%%%%%%%%%%%%%%%%%%%%%%%%%%%%%%%%%%%%%%%%%%%%%%%%%%%

\end{document}